\documentclass[aps,prl,showpacs,twocolumn,superscriptaddress,nofootinbib]{revtex4-1}  
\usepackage{graphicx,color}  
\usepackage{dcolumn}   
\usepackage{bm}        
\usepackage{amssymb} 
\usepackage{amsmath}
 \newcommand{\beq}{\begin{equation}}
\newcommand{\eeq}{\end{equation}}
\newcommand{\beqa}{\begin{eqnarray}}
\newcommand{\eeqa}{\end{eqnarray}}
\newcommand{\be}{\begin{equation}}
\newcommand{\ee}{\end{equation}}
\newcommand{\bea}{\begin{eqnarray}}
\newcommand{\eea}{\end{eqnarray}}

\begin{document}
\def\eg{{\it e.g.}}
\newcommand{\nc}{\newcommand}
\nc{\rnc}{\renewcommand}
\rnc{\d}{\mathrm{d}}
\nc{\D}{\partial}
\nc{\K}{\kappa}
\nc{\bK}{\bar{\K}}
\nc{\bN}{\bar{N}}
\nc{\bq}{\bar{q}}
\nc{\vbq}{\vec{\bar{q}}}
\nc{\g}{\gamma}
\nc{\lrarrow}{\leftrightarrow}
\nc{\rg}{\sqrt{g}}
\rnc{\[}{\begin{equation}}
\rnc{\]}{\end{equation}}
\nc{\nn}{\nonumber}
\rnc{\(}{\left(}
\rnc{\)}{\right)}
\nc{\q}{\vec{q}}
\nc{\x}{\vec{x}}
\rnc{\a}{\hat{a}}
\nc{\ep}{\epsilon}
\nc{\tto}{\rightarrow}
\rnc{\inf}{\infty}
\rnc{\Re}{\mathrm{Re}}
\rnc{\Im}{\mathrm{Im}}
\nc{\z}{\zeta}
\nc{\mA}{\mathcal{A}}
\nc{\mB}{\mathcal{B}}
\nc{\mC}{\mathcal{C}}
\nc{\mD}{\mathcal{D}}
\nc{\mN}{\mathcal{N}}
\rnc{\H}{\mathcal{H}}
\rnc{\L}{\mathcal{L}}
\nc{\<}{\langle}
\rnc{\>}{\rangle}
\nc{\fnl}{f_{NL}}
\nc{\gnl}{g_{NL}}
\nc{\fnleq}{f_{NL}^{equil.}}
\nc{\fnlloc}{f_{NL}^{local}}
\nc{\vphi}{\varphi}
\nc{\Lie}{\pounds}
\nc{\half}{\frac{1}{2}}
\nc{\bOmega}{\bar{\Omega}}
\nc{\bLambda}{\bar{\Lambda}}
\nc{\dN}{\delta N}
\nc{\gYM}{g_{\mathrm{YM}}}
\nc{\geff}{g_{\mathrm{eff}}}
\nc{\tr}{\mathrm{tr}}
\nc{\oa}{\stackrel{\leftrightarrow}}
\nc{\IR}{{\rm IR}}
\nc{\UV}{{\rm UV}}

\title{Einstein Gauss-Bonnet Theories  \\
 as Ordinary, Wess-Zumino Conformal Anomaly Actions} 

\author{Claudio Corian\`o}
\affiliation{Dipartimento di Matematica e Fisica, Universit\`{a} del Salento \\ and \\ INFN Sezione di Lecce, Via Arnesano, 73100 Lecce, Italy} 
\author{Matteo Maria Maglio}\affiliation{Galileo Galilei Institute for Theoretical Physics, \\
	Largo Enrico Fermi 2, I-50125 Firenze, Italy}\affiliation{Institute for Theoretical Physics (ITP), University of Heidelberg\\
	Philosophenweg 16, 69120 Heidelberg, Germany}

\date{\today}

\begin{abstract}
Recently, the possibility of evading Lovelock's theorem at $d=4$, via a singular redefinition of the dimensionless coupling of the Gauss-Bonnet term, has been very extensively discussed in the cosmological context. The term is added as a quadratic contribution of the curvature tensor to the Einstein-Hilbert action, originating theories of  "Einstein Gauss-Bonnet" (EGB) type. 
We point out that the action obtained by the dimensional regularization procedure, implemented with the extraction of a single conformal factor, correspond just to an ordinary Wess-Zumino anomaly action, even though it is deprived of the contribution from the Weyl tensor. 
We also show that a purely gravitational version of the EGB theory can be generated by allowing a finite renormalization of the Gauss-Bonnet topological contribution at $d= 4 + \epsilon$, as pointed out by Mazur and Mottola. The result is an effective action which is quadratic, rather then quartic, in the dilaton field, and scale free, compared to the previous derivations. 
The dilaton, in this case can be removed from the spectrum, leaving a pure gravitational theory, which is nonlocal. We comment on the physical meaning of the two types of actions, 
which may be used to describe such topological terms both below and above the conformal breaking scale.
\end{abstract}

\maketitle

\section{Introduction}
Corrections to the Einstein-Hilbert action, in the form of higher powers of the Riemann curvature $R_{\mu\nu\alpha\beta}$ and of its contractions, have been extensively discussed in the cosmological context. The goal of such modifications is to deepen our insight into the problem of the dark energy dominance in the evolution of our universe, providing an answer to the hierarchy problem, implicit in the current fits of the cosmological constant, within the $\Lambda$CDM ($\Lambda$-Cold Dark Matter) model \cite{Antoniadis:2011ib}. \\
Higher derivative gravity theories have to face, in general, fundamental consistency issues, due to the presence of ghost states in their spectra, at quantum level. 
Dimension-4 corrections, quadratic in the Riemann tensor, may appear in the form of Weyl-invariant densities, multiplied by dimensionless couplings. At specific spacetime dimensions, some of these invariants may generate equations of motion of the second order, and are classified by Lovelock's theorem \cite{Lovelock:1971yv}(see \cite{Charmousis:2014mia}). The same theorem guarantees, under ordinary conditions, such as smoothness of the action and locality of the interactions, that the Einstein Hilbert action with a cosmological constant $\Lambda$
\beq
\mathcal{S}_{EH}=\int d^d x \sqrt{g}(M_P^2 R + \Lambda)  
\eeq 
is the only one that provides equations of motion of the second order at $d=4$. At $d=2$, the EH action 
(with the Planck mass $M_P\to 1$) is topological, and shares features similar to other topological terms 
($E_4, E_6$ etc) quadratic and cubic in $R_{\mu\nu\alpha\beta}$ as, for example,  in $d=4$ and  6.
Recently, the proposal of a modified theory of gravity \cite{Glavan:2019inb} based on the addition of a Gauss-Bonnet (GB) term $V_E(d=4)$ to the EH action, has received very significant attention \cite{Hennigar:2020lsl,Fernandes:2020nbq,Easson:2020mpq,Kobayashi:2020wqy,Konoplya:2020qqh,Bonifacio:2020vbk,Ai:2020peo,Wei:2020ght,Aoki:2020lig,Nojiri:2020tph,Konoplya:2020bxa,Guo:2020zmf,Fernandes:2020rpa,Casalino:2020kbt,Hegde:2020xlv,Ghosh:2020vpc,Doneva:2020ped,Zhang:2020qew,Konoplya:2020ibi,Singh:2020xju,Ghosh:2020syx,Konoplya:2020juj,Kumar:2020uyz,Zhang:2020qam,HosseiniMansoori:2020yfj,Wei:2020poh,Singh:2020nwo,Churilova:2020aca,Islam:2020xmy,Mishra:2020gce,Konoplya:2020cbv,Zhang:2020sjh,EslamPanah:2020hoj,Aragon:2020qdc,Aoki:2020iwm,Shu:2020cjw,Mahapatra:2020rds,Lu:2020iav,Gurses:2020ofy,Banerjee:2020dad,Ge:2020tid,Yang:2020jno,Lin:2020kqe,Yang:2020czk}. This takes the form

\beq
\label{GB0}
\alpha V_E(d)=\alpha \int d^d x \sqrt{g}E, 
\eeq
where $E\equiv E_4$ is the Euler-Poincar\`e topological density

\beqa
\label{GB1}
 E& =& R^2 - 4 R^{\mu \nu} R_{\mu \nu} + R^{\mu \nu \rho \sigma} R_{\mu \nu \rho \sigma}. 
\eeqa
$V_E$, as just mentioned, is added to the Einstein-Hilbert action (EH), 
\beq
\mathcal{S}_{EGB}=S_{EH} + \alpha V_E,
\label{first}
\eeq

and $\alpha$ is a dimensionless coupling constant. The addition of this term, in general,  
generates equations of motion of the form 
 \beq
\frac{1}{\kappa}\left(R_{\mu\nu}-\frac{1}{2}g_{\mu\nu}R+\Lambda_{0}g_{\mu\nu}\right)+\alpha (V_{E}(d))_{\mu\nu}=0,
\label{GB2}
\eeq
with $(V_{E}(d))_{\mu\nu}={\delta V_E(d)}/{\delta g_{\mu\nu}}$.
However, due to the topological nature of \eqref{GB0} at $d=4$, the variation ($\delta_g$) of the integrand 
in \eqref{GB0} with respect to the metric, is an ordinary boundary contribution

\beq
\delta_g (\sqrt{g} E)=\sqrt{g}\,\nabla_\sigma \delta X^\sigma,
\delta X^\sigma=\varepsilon^{\mu\nu\alpha \beta}\varepsilon^{\sigma\lambda\gamma\tau}
\delta_g \Gamma^\eta_{\nu\lambda}g_{\mu\eta}R_{\alpha \beta \gamma \tau}, 
\eeq
where $\varepsilon^{\mu\nu\alpha \beta}={\epsilon^{\mu\nu\alpha \beta}}/{\sqrt{g}}$, and $\Gamma$ is the Christoffel connection in the metric $g_{\mu\nu}$.  
$(V_E(d))^{\mu\nu}$ is therefore classified as an evanescent term at $d=4$
\beqa
(V_E(4))^{\mu\nu}&=& 4R_{\mu\alpha\beta\sigma}R^{\;\,\alpha\beta\sigma}_\nu-8R_{\mu\alpha\nu\beta}R^{\alpha\beta}\nonumber \\
&& -8R_{\mu\alpha}R^{\;\,\alpha}_{\nu}+4RR_{\mu\nu}-g_{\mu\nu}{E}=0,    \qquad (d=4)\nn\\
\eeqa
if we assume asymptotic flatness and strictly $d=4$. \\
In \cite{Glavan:2019inb} it was observed that it is possible to 
renormalize the coupling $\alpha$ in \eqref{GB2} by the replacement 
$\alpha\to \alpha/(d-4)$, and perform the $d\to 4$ limit on the equation, generating a 0/0 term in such modified version of \eqref{GB2}, which may yield finite equations of motion. The limit, however, is 
non-trivial, and extra degrees of freedom are introduced in this procedure. They are either derived from the extra dimensional components of the metric, or from the choice of a particular parameterization, involving the extraction of a conformal factor.\\
\section{A reapprisal}
Naively, the approach of \cite{Glavan:2019inb} generates a limiting theory with dynamical equations of motion of the second order. However, for the procedure to be consistent, it has to be correctly framed in the context of dimensional regularization (DR), since it violates conservation of the stress energy tensor \cite{Gurses:2020rxb}. \\
We are going to show that some subtle issues, related to the possibility of performing an {\em additional} finite renormalizations of $V_E$,  not addressed in the previous literature on EGBs, should be also taken into account. The result of our analysis is the proposal of a nonlocal EGB theory, from which the dilaton field is classically integrated out. It stands on its own as an alternative realization of such theories, and it is purely gravitational.  At the same time, we will present a consistent interpretation of the actions derived so far in the context of graviton-dilaton theories in \cite{Fernandes:2020nbq,Hennigar:2020lsl,Lu:2020iav}, and in other related papers on the subject.  \\
Notice that EGB theories of the type discussed in \cite{Glavan:2019inb} \cite{Fernandes:2020nbq,Hennigar:2020lsl,Lu:2020iav} are purely classical, even though the GB term present in their definition takes the typical form of an ordinary "countertem" in DR. Given 
the evanescent character of $V_E(d)$ at $d=4$, its inclusion at quantum level is not necessary, in a strict sense, to eliminate the UV divergences of such theories. Rather, the term emerges by imposing that the renormalized (quantum) effective action, obtained by integrating in the partition function a certain conformal sector \cite{Coriano:2021nvn}, satisfies the WZ consistency condition. \\

\section{Conformal decompositions and Wess-Zumino actions}

If we extend the integration region of $V_E$ from 4 to $d$ as in \eqref{GB0},  the corresponding limiting action 

\beq
\label{bonc}
\mathcal{S}_{EGB} =\mathcal{S}_{EH} +\mathcal{S}_{GB}(d) \qquad \mathcal{S}_{GB}(d)=\frac{\alpha}{\epsilon}V_E(d)
\eeq
naively appears to be finite at $d=4$, and modifies \eqref{GB0}. It is purely gravitational and quadratic in $R_{\mu\nu\rho\sigma}$. As such, 
the EGB theory appears to violate Lovelock's theorem. 
Some important criticism to the procedure in \cite{Glavan:2019inb} has been raised by several parts. In \cite{Gurses:2020rxb} the authors have pointed out the difficulty of the tensorial limit of $V_E^{\mu\nu}$ from generic $d$ to $d=4$, while in \cite{Fernandes:2020nbq,Hennigar:2020lsl,Lu:2020iav} a consistent regularization of the theory 
\eqref{bonc} has been proposed.
Such regularizations identify the regulated action as of WZ type, as we are going to show, for being 
defined via a procedure of Weyl-gauging. The approach consists in introducing a fiducial metric $\bar{g}_{\mu\nu}$, in the form 
$g_{\mu\nu}=\bar{g}_{\mu\nu} e^{2\phi}$, where we extract the conformal factor via a dilaton field $\phi$. Notice that this separation is invariant under the symmetry 
\beq
\label{sym}
\phi\to \phi -\sigma , \,\,\bar{g}_{\mu\nu}\to \bar{g}_{\mu\nu} e^{2\sigma}
\eeq
which constraints the equations of motion of $\bar{g}_{\mu\nu}$ and $\phi$, and 
explains the relation between the two equations of motion of these two fields, noticed in \cite{Fernandes:2020nbq,Hennigar:2020lsl}. The WZ action of the GB term $V_E(g_{\mu\nu}, d)$ is defined as the difference 
\beq
\mathcal{S}^{(WZ)}_{GB}\equiv\frac{\alpha}{\epsilon}\left(V_E(\bar{g}_{\mu\nu}e^{2\phi},d)- V_E(\bar{g}_{\mu\nu},d)\right)
\label{ss}
\eeq
where 
\beq
\frac{1}{\epsilon}\frac{\delta V_E(\bar{g}_{\mu\nu}e^{2\phi},d)}{\delta\phi}=\sqrt{g}E(g)
\eeq
while the expansion in $\epsilon$ around $d=4$ for $V_E(\bar{g}_{\mu\nu}e^{2\phi},d)$
\beqa 
\frac{1}{\epsilon}V_E(\bar{g}_{\mu\nu}e^{2\phi},d)&=&\frac{1}{\epsilon}V_E(\bar{g}_{\mu\nu},d=4) + 
V'_E(\bar{g}_{\mu\nu},\phi, d=4)\nonumber \\
\label{exp}
\eeqa
and a similar one for  $V_E(\bar{g}_{\mu\nu},d)$, in the limit $\epsilon\to 0$,  combined 
give 
\beq
\mathcal{S}^{(WZ)}_{GB}=\alpha V'_E(\bar{g}_{\mu\nu},\phi, d=4), 
\label{ww}
\eeq
having dropped a Weyl-invariant terms $V'_E(\bar{g}_{\mu\nu},d=4)$ in the dimensional expansion around $d=4$. The dimensional reduction that takes to \eqref{ww} requires some care, due to the presence of cutoffs in the extra dimensions and will be discussed elsewhere \cite{CMT}. Neglecting some Weyl-invariant terms, $V'_E$ 
can be given in the form derived in \cite{Fernandes:2020nbq,Hennigar:2020lsl,Lu:2020iav} 
\beqa
\mathcal{S}^{(WZ)}_{GB}&=&
\alpha\int d^4 x \sqrt{\bar{g}} \biggl[\phi \bar{E}-\Big(4 \bar{G}^{\mu\nu}(\bar{\nabla}_\mu\phi\bar{\nabla}_\nu\phi)\nn\\
 &&+2(\bar{\nabla}_\lambda \phi\bar{\nabla}^\lambda \phi)^2
 +4\bar\Box\phi\bar{\nabla}_\lambda \phi\bar{\nabla}^\lambda \phi\Big)\biggl],
 \label{GGB}
\eeqa
where $\bar{G}_{\mu\nu}$ is the Einstein tensor in the fiducial metric $\bar{g}_{\mu\nu}$. Counterexamples to such  procedure, where the action is not properly regulated, are those in which the $d-4$ extra space components of $\bar{g}_{\mu\nu}$ are flat, as discussed in \cite{Gurses:2020rxb}. Such cases need to be excluded, for consistency \cite{CMT}.  
  
Notice that the equations of motion for the metric, from   
$\bar{g}_{\mu\nu}$ and $\phi$ are constrained by the relations 
\beqa
\label{fin}
\left(2 {g}_{\mu\nu}\frac{\delta}{\delta _{\mu\nu}}-\frac{\delta}{\delta\phi}\right)\mathcal{S}^{(WZ)}_{GB}=-\alpha\sqrt{\bar g}\bar{E},\nn\\
\eeqa
\beqa
&&2{g}_{\mu\nu}\frac{\delta \mathcal{S}^{(WZ)}_{GB}}{\delta {g}_{\mu\nu}}=\sqrt{g}E - 
\sqrt{\bar g}\bar E, 
\label{equal}
\eeqa
\beq
\frac{\delta \mathcal{S}^{(WZ)}_{GB}}{\delta \phi}=\sqrt{g}E,
\eeq
similarly to the case of anomaly actions, but with the omission of the Weyl tensor squared term. 
\section{Conformal Symmetry breaking}
WZ actions have been identified in the past either by the Weyl gauging procedure, as shown above, or, equivalently, by the Noether "leaking" method. Both methods stop at order $d$ in the dilaton field, if we are in a $d$-dimensional spacetime. The actions obtained are local, since they include a dilaton field. If we were able to solve for $\phi$, then the actions would become nonlocal, in agreement with the general lore that anomalies are not related to local actions. \\
The separation of the conformal factor from the fiducial 
metric has necessarily to be associated with the breaking of the conformal symmetry.  
  In these formulations, the dilaton $\phi$ can either be part of the regulated action, as discussed in \cite{Fernandes:2020nbq,Hennigar:2020lsl,Lu:2020iav}, or can be eliminated, by solving for this field in terms of the background metric.\\
   In the Lagrangians of the first class 
 (dilaton gravities), $\phi$ is generally interpreted as a physical particle. In this case it is convenient to introduce the parameterization 
 \beq
 e^\phi=1- \frac{\chi}{f}
 \eeq
 where $f$ is a conformal breaking scale,
 which breaks the symmetry in both the first and the second term of \eqref{ss}. In this case, the action can be expanded around the field value $\phi=\phi_0=0$ (i.e. $ \chi=\chi_0=0$), but it is clear that the selection of such values for $\phi$ or, equivalently,  $\chi$, should occur via an extra potential which is clearly not included in $S^{(GB)}_{WZ}$. A mechanism of spontaneous breaking of the local conformal symmetry needs to be invoked, after the inclusion of an extra potential. As pointed out in  \cite{tHooft:2016uxd,tHooft:2015koi} this spontaneous breaking of the symmetry and its restoration may be one one of the most important issue in particle theory.
 Notice, however, that $S_{WZ}$ already breaks the symmetry \eqref{ss} by the anomaly since 
 \beq
 \frac{\delta S^{(GB)}_{WZ}}{\delta \sigma}=-\sqrt{\bar g}\bar{E}. 
 \eeq
 However, this breaking is expected to be part of a wider picture not captured by \eqref{ss}.  \\
 In general, theories of this type, characterised by the presence of a Nambu-Goldstone local shifting mode, share a similar behaviour, another example being that of Stuckelberg models, where the extra potential is assumed to be of instanton origin, and mixes an ordinary Higgs sector with the Stuckleberg field  \cite{Coriano:2005js}.  

 Actions belonging to this class, quartic in the field $\phi$, obviously depend on the choice of the form of 
 $\bar{g}_{\mu\nu}$ in DR and  may differ, therefore,  by finite $\phi$-dependent terms. \\
 This is well-known in the case of ordinary Kaluza-Klein theories, whose Lagrangians depend on the manifold of compactification. In the presence of additional scales in the extra dimensions (an example are extra manifolds maximally symmetric such as Einstein spaces), the resulting actions do not preserve the residual Weyl symmetry \eqref{sym}, but are only invariant under global dilatations \cite{CMT}.   
This points, obviously, towards their non-uniqueness, since the options are manyfolds.\\
From our perspective, such actions are useful in the description of the behaviour of a Weyl-invariant theory below the scale $f$, but are unsuitable to describe the effects of the anomaly, which is part of these regularizations, at larger scales, where a dimensionful expansion parameter should be absent. Perturbations around flat space indicate that the role of such "parameter" is the nonlocal term $R\Box^{-1}$, as predicted from an in-depth analysis of the conformal Ward identities statisfied by the quantum effective action \cite{Coriano:2021nvn} \cite{Coriano:2017mux}. 

As we are going to show, also in the EGB case, a specific finite renormalization of $V_E$ generates an action which is only quadratic in $\phi$, and may capture consistently the  the UV behaviour of the anomaly. It can be written in a nonlocal form. In this case the expansion of this second type of actions is in terms of $R\Box^{-1}$, as discussed in  \cite{Coriano:2021nvn} \cite{Coriano:2017mux}. Also in this case, the identification of this action takes is not unique, since it is identified by a Weyl scaling procedure that is insensitive to Weyl-invariant contributions.

\section{Nonlocal EGB theories}
As already mentioned, nonlocal versions of such actions are also possible. 
A modification of the GB term at $O(\epsilon)$, generates anomaly actions of a different character compared with the previous ones \cite{Fernandes:2020nbq,Hennigar:2020lsl,Lu:2020iav}. Indeed, we have the possibility of performing a finite renormalization of the GB term in DR, as already discussed in \cite{Mazur:2001aa} in the context of conformal anomaly actions, to end up with a very different theory.

For this reason, one can simply modify by a finite  renormalization  the structure of the GB term away from $d=4$ in the form
 \beq
E_{ext}=E + \frac{\epsilon}{2 (d-1)^2} R^2,\qquad \tilde{V}_E=\int d^d x \sqrt{g}E_{ext}. 
 \eeq
 with $V_E\to \tilde{V}_E$ in \eqref{ss}, obtaining

\begin{equation}
\mathcal{S}^{(WZ)}_{GB} =\frac{\alpha}{\epsilon}\left(\tilde{V}_E(\bar{g}_{\mu\nu}e^{2\phi},d)- \tilde{V}_E(\bar{g}_{\mu\nu},d\right).
\label{inter}
\eeq
Performing the $d\to 4$ limit and using an expansion similar  to \eqref{exp}, one derives the regulated GB action
\begin{equation}
\mathcal{S}^{(WZ)}_{GB} = \alpha\int\,d^4x\,\sqrt{-\bar g}\,\left\{\left(\overline E_4 - {2\over 3}
\bar{\Box} \overline R\right)\phi + 2\,\phi\bar\Delta_4\phi\right\},\,
\label{WZ2}
\end{equation}
where $\bar{\Delta}_4$ is the quartic conformally covariant operator 
\cite{Riegert:1987kt}
 \begin{equation}
\Delta_4 \equiv \Box^2 + 2 R^{\mu\nu}\nabla_\mu\nabla_\nu +{1\over 3} (\nabla^\mu R)
\nabla_\mu - {2\over 3} R \Box,
\label{Delta}
\end{equation}
evaluated with respect to the fiducial metric $\bar{g}_{\mu\nu}$. \\
Eq. \eqref{WZ2} shows that it is possible to define a quadratic - and not only a quartic -  action in $\phi$, for the regularized GB term. Combined with the EH action, this would define a version of the EGB theory which is purely gravitational, since $\phi$ can be eliminated. To derive its nonlocal expression, we can use the relation
\beq
\frac{\delta}{\delta\phi}\frac{1}{\epsilon}\tilde{V}_E(g_{\mu\nu},d)= \sqrt{g}\left(E-\frac{2}{3}\Box R +
\epsilon\frac{R^2}{2(d-1)^2}\right)
\eeq
in \eqref{inter}, giving  
 \beqa
 \frac{\delta \mathcal{S}^{(WZ)}_{GB}}{\delta\phi}&=&\alpha\sqrt{g}\left(E-\frac{2}{3}\Box R \right)\nonumber \\
&=&\alpha\sqrt{\bar g}\left(\bar E-\frac{2}{3}\bar \Box\bar R + 4 \bar\Delta_4 \phi\right).
\label{solve}
\eeqa
Using the relation $\sqrt{\bar{g}}\,\bar{\Delta}_4= \sqrt{g}\,\Delta_4 $, valid on conformal scalars, 
and introducing the Green function $D_4(x,y)$ of $\Delta_4$
\beq
\sqrt{g}\,\Delta_4 D_4(x,y)=\delta^4(x,y),
\eeq
we can solve for $\phi$ in \eqref{solve}. Inserting its expression back into \eqref{WZ2} 
we obtain the nonlocal action
\beqa
 \mathcal{S}^{(WZ)}_{GB}& =& 
{\alpha\over 8} \int d^4x\,\sqrt{-g}\, \int d^4x'\,\sqrt{-g'}\,
\left(E_4 - {2\over 3} \Box R\right)_x\, \nonumber \\
&&\qquad \times D_4(x,x')\left(E_4- {2\over 3} \Box R\right)_{x'},\,
\label{anomact}
\eeqa
that coincides with the result provided in \cite{Mazur:2001aa} by Mazur and Mottola. \\
A nonlocal EGB theory of gravity follows quite directly from this action, once we add to \eqref{anomact} 
the EH term.  As clear from \eqref{solve}, the dilaton $\phi$ is removed by solving that equation by the Green function of $\Delta_4$. The operator is quartic, and as such, at a first look, it seems to defeat any possibility to describe a theory of second order. However, as discussed in recent analysis, the functional expansion of nonlocal actions of this type in the background metric - its vertices - can be organized in terms of operatorial insertions of the form $R\Box^{-1}$ for each external leg, at least around flat space \cite{Coriano:2017mux,Coriano:2021nvn}.
Obviously, the scaling approach followed in order to remove the dilaton,  is unable to identify possible Weyl-invariant (i.e. $\phi$-independent ) contributions, functions only of the background metric $\bar{g}_{\mu\nu}$. 

\section{ Conclusions}
Topological terms in gravity, introduced in a purely classical context, share a significant overlap with former analysis of the anomaly actions of Wess-Zumino forms, widely discussed in the related literature. In the case of the $d=4$ GB term, its addition to the Einstein term at $d=4$, had tentatively opened the way to a new form of gravity, EGB gravity. This has sparked a wide interest for 
possibly evading, via a singular limit, Lovelock's theorem.  Subsequent analysis showed that such theories should be correctly interpreted as of Horndeski type, in the form of dilaton gravities. We have pointed out that such class of actions are long-known, specific forms of WZ dilaton actions. 
We have also shown that a finite renormalization of the GB  density, which is perfectly allowed in DR, gives the possibility of generating an EGB theory which is nonlocal. What we consider important is that such two classes of 
actions are suitable for describing the role of topological terms in two different domains, the IR and the UV, using, respectively, either local or a nonlocal formulation. Nonlocal EGB theories appears to be deprived of double poles at least up to 4th order in the fluctuations of the metric, in an expansion around flat space, and can generate perturbations of the form $f(R\, \square^{-1})$, as shown in recent analysis.   

\centerline{\bf Acknowledgements }
We thank Dimosthenis Theofilopoulos for collaborating to related work and to Mario Cret\`i, Riccardo Tommasi and Stefano Lionetti for discussions. We thank Pietro Colangelo and Paul Frampton for discussions and for a reading of the manuscript.
We dedicate this work to the memory of our Colleague and friend Pierluigi Santo. This work is partially supported by INFN under Iniziativa Specifica QFT-HEP.
M. M. M. is supported by the European Research Council (ERC) under the European Union’s Horizon 2020 research and innovation program (grant agreement No818066) and by Deutsche Forschungsgemeinschaft (DFG, German Research Foundation) under Germany's Excellence Strategy EXC-2181/1 - 390900948 (the Heidelberg STRUCTURES Cluster of Excellence).

%


\begin{thebibliography}{57}%
\makeatletter
\providecommand \@ifxundefined [1]{%
 \@ifx{#1\undefined}
}%
\providecommand \@ifnum [1]{%
 \ifnum #1\expandafter \@firstoftwo
 \else \expandafter \@secondoftwo
 \fi
}%
\providecommand \@ifx [1]{%
 \ifx #1\expandafter \@firstoftwo
 \else \expandafter \@secondoftwo
 \fi
}%
\providecommand \natexlab [1]{#1}%
\providecommand \enquote  [1]{``#1''}%
\providecommand \bibnamefont  [1]{#1}%
\providecommand \bibfnamefont [1]{#1}%
\providecommand \citenamefont [1]{#1}%
\providecommand \href@noop [0]{\@secondoftwo}%
\providecommand \href [0]{\begingroup \@sanitize@url \@href}%
\providecommand \@href[1]{\@@startlink{#1}\@@href}%
\providecommand \@@href[1]{\endgroup#1\@@endlink}%
\providecommand \@sanitize@url [0]{\catcode `\\12\catcode `\$12\catcode
  `\&12\catcode `\#12\catcode `\^12\catcode `\_12\catcode `\%12\relax}%
\providecommand \@@startlink[1]{}%
\providecommand \@@endlink[0]{}%
\providecommand \url  [0]{\begingroup\@sanitize@url \@url }%
\providecommand \@url [1]{\endgroup\@href {#1}{\urlprefix }}%
\providecommand \urlprefix  [0]{URL }%
\providecommand \Eprint [0]{\href }%
\providecommand \doibase [0]{http://dx.doi.org/}%
\providecommand \selectlanguage [0]{\@gobble}%
\providecommand \bibinfo  [0]{\@secondoftwo}%
\providecommand \bibfield  [0]{\@secondoftwo}%
\providecommand \translation [1]{[#1]}%
\providecommand \BibitemOpen [0]{}%
\providecommand \bibitemStop [0]{}%
\providecommand \bibitemNoStop [0]{.\EOS\space}%
\providecommand \EOS [0]{\spacefactor3000\relax}%
\providecommand \BibitemShut  [1]{\csname bibitem#1\endcsname}%
\let\auto@bib@innerbib\@empty
\bibitem [{\citenamefont {Antoniadis}\ \emph {et~al.}(2012)\citenamefont
  {Antoniadis}, \citenamefont {Mazur},\ and\ \citenamefont
  {Mottola}}]{Antoniadis:2011ib}%
  \BibitemOpen
  \bibfield  {author} {\bibinfo {author} {\bibfnamefont {I.}~\bibnamefont
  {Antoniadis}}, \bibinfo {author} {\bibfnamefont {P.~O.}\ \bibnamefont
  {Mazur}}, \ and\ \bibinfo {author} {\bibfnamefont {E.}~\bibnamefont
  {Mottola}},\ }\href {\doibase 10.1088/1475-7516/2012/09/024} {\bibfield
  {journal} {\bibinfo  {journal} {JCAP}\ }\textbf {\bibinfo {volume} {1209}},\
  \bibinfo {pages} {024} (\bibinfo {year} {2012})},\ \Eprint
  {http://arxiv.org/abs/1103.4164} {arXiv:1103.4164 [gr-qc]} \BibitemShut
  {NoStop}%
\bibitem [{\citenamefont {Lovelock}(1971)}]{Lovelock:1971yv}%
  \BibitemOpen
  \bibfield  {author} {\bibinfo {author} {\bibfnamefont {D.}~\bibnamefont
  {Lovelock}},\ }\href {\doibase 10.1063/1.1665613} {\bibfield  {journal}
  {\bibinfo  {journal} {J. Math. Phys.}\ }\textbf {\bibinfo {volume} {12}},\
  \bibinfo {pages} {498} (\bibinfo {year} {1971})}\BibitemShut {NoStop}%
\bibitem [{\citenamefont {Charmousis}(2015)}]{Charmousis:2014mia}%
  \BibitemOpen
  \bibfield  {author} {\bibinfo {author} {\bibfnamefont {C.}~\bibnamefont
  {Charmousis}},\ }\href {\doibase 10.1007/978-3-319-10070-8_2} {\bibfield
  {journal} {\bibinfo  {journal} {Lect. Notes Phys.}\ }\textbf {\bibinfo
  {volume} {892}},\ \bibinfo {pages} {25} (\bibinfo {year} {2015})},\ \Eprint
  {http://arxiv.org/abs/1405.1612} {arXiv:1405.1612 [gr-qc]} \BibitemShut
  {NoStop}%
\bibitem [{\citenamefont {Glavan}\ and\ \citenamefont
  {Lin}(2020)}]{Glavan:2019inb}%
  \BibitemOpen
  \bibfield  {author} {\bibinfo {author} {\bibfnamefont {D.}~\bibnamefont
  {Glavan}}\ and\ \bibinfo {author} {\bibfnamefont {C.}~\bibnamefont {Lin}},\
  }\href {\doibase 10.1103/PhysRevLett.124.081301} {\bibfield  {journal}
  {\bibinfo  {journal} {Phys. Rev. Lett.}\ }\textbf {\bibinfo {volume} {124}},\
  \bibinfo {pages} {081301} (\bibinfo {year} {2020})},\ \Eprint
  {http://arxiv.org/abs/1905.03601} {arXiv:1905.03601 [gr-qc]} \BibitemShut
  {NoStop}%
\bibitem [{\citenamefont {Hennigar}\ \emph {et~al.}(2020)\citenamefont
  {Hennigar}, \citenamefont {Kubiz\v{n}\'ak}, \citenamefont {Mann},\ and\
  \citenamefont {Pollack}}]{Hennigar:2020lsl}%
  \BibitemOpen
  \bibfield  {author} {\bibinfo {author} {\bibfnamefont {R.~A.}\ \bibnamefont
  {Hennigar}}, \bibinfo {author} {\bibfnamefont {D.}~\bibnamefont
  {Kubiz\v{n}\'ak}}, \bibinfo {author} {\bibfnamefont {R.~B.}\ \bibnamefont
  {Mann}}, \ and\ \bibinfo {author} {\bibfnamefont {C.}~\bibnamefont
  {Pollack}},\ }\href {\doibase 10.1007/JHEP07(2020)027} {\bibfield  {journal}
  {\bibinfo  {journal} {JHEP}\ }\textbf {\bibinfo {volume} {07}},\ \bibinfo
  {pages} {027} (\bibinfo {year} {2020})},\ \Eprint
  {http://arxiv.org/abs/2004.09472} {arXiv:2004.09472 [gr-qc]} \BibitemShut
  {NoStop}%
\bibitem [{\citenamefont {Fernandes}\ \emph {et~al.}(2020)\citenamefont
  {Fernandes}, \citenamefont {Carrilho}, \citenamefont {Clifton},\ and\
  \citenamefont {Mulryne}}]{Fernandes:2020nbq}%
  \BibitemOpen
  \bibfield  {author} {\bibinfo {author} {\bibfnamefont {P.~G.~S.}\
  \bibnamefont {Fernandes}}, \bibinfo {author} {\bibfnamefont {P.}~\bibnamefont
  {Carrilho}}, \bibinfo {author} {\bibfnamefont {T.}~\bibnamefont {Clifton}}, \
  and\ \bibinfo {author} {\bibfnamefont {D.~J.}\ \bibnamefont {Mulryne}},\
  }\href {\doibase 10.1103/PhysRevD.102.024025} {\bibfield  {journal} {\bibinfo
   {journal} {Phys. Rev. D}\ }\textbf {\bibinfo {volume} {102}},\ \bibinfo
  {pages} {024025} (\bibinfo {year} {2020})},\ \Eprint
  {http://arxiv.org/abs/2004.08362} {arXiv:2004.08362 [gr-qc]} \BibitemShut
  {NoStop}%
\bibitem [{\citenamefont {Easson}\ \emph {et~al.}(2020)\citenamefont {Easson},
  \citenamefont {Manton},\ and\ \citenamefont {Svesko}}]{Easson:2020mpq}%
  \BibitemOpen
  \bibfield  {author} {\bibinfo {author} {\bibfnamefont {D.~A.}\ \bibnamefont
  {Easson}}, \bibinfo {author} {\bibfnamefont {T.}~\bibnamefont {Manton}}, \
  and\ \bibinfo {author} {\bibfnamefont {A.}~\bibnamefont {Svesko}},\ }\href
  {\doibase 10.1088/1475-7516/2020/10/026} {\bibfield  {journal} {\bibinfo
  {journal} {JCAP}\ }\textbf {\bibinfo {volume} {10}},\ \bibinfo {pages} {026}
  (\bibinfo {year} {2020})},\ \Eprint {http://arxiv.org/abs/2005.12292}
  {arXiv:2005.12292 [hep-th]} \BibitemShut {NoStop}%
\bibitem [{\citenamefont {Kobayashi}(2020)}]{Kobayashi:2020wqy}%
  \BibitemOpen
  \bibfield  {author} {\bibinfo {author} {\bibfnamefont {T.}~\bibnamefont
  {Kobayashi}},\ }\href {\doibase 10.1088/1475-7516/2020/07/013} {\bibfield
  {journal} {\bibinfo  {journal} {JCAP}\ }\textbf {\bibinfo {volume} {07}},\
  \bibinfo {pages} {013} (\bibinfo {year} {2020})},\ \Eprint
  {http://arxiv.org/abs/2003.12771} {arXiv:2003.12771 [gr-qc]} \BibitemShut
  {NoStop}%
\bibitem [{\citenamefont {Konoplya}\ and\ \citenamefont
  {Zhidenko}(2020{\natexlab{a}})}]{Konoplya:2020qqh}%
  \BibitemOpen
  \bibfield  {author} {\bibinfo {author} {\bibfnamefont {R.~A.}\ \bibnamefont
  {Konoplya}}\ and\ \bibinfo {author} {\bibfnamefont {A.}~\bibnamefont
  {Zhidenko}},\ }\href {\doibase 10.1103/PhysRevD.101.084038} {\bibfield
  {journal} {\bibinfo  {journal} {Phys. Rev. D}\ }\textbf {\bibinfo {volume}
  {101}},\ \bibinfo {pages} {084038} (\bibinfo {year} {2020}{\natexlab{a}})},\
  \Eprint {http://arxiv.org/abs/2003.07788} {arXiv:2003.07788 [gr-qc]}
  \BibitemShut {NoStop}%
\bibitem [{\citenamefont {Bonifacio}\ \emph {et~al.}(2020)\citenamefont
  {Bonifacio}, \citenamefont {Hinterbichler},\ and\ \citenamefont
  {Johnson}}]{Bonifacio:2020vbk}%
  \BibitemOpen
  \bibfield  {author} {\bibinfo {author} {\bibfnamefont {J.}~\bibnamefont
  {Bonifacio}}, \bibinfo {author} {\bibfnamefont {K.}~\bibnamefont
  {Hinterbichler}}, \ and\ \bibinfo {author} {\bibfnamefont {L.~A.}\
  \bibnamefont {Johnson}},\ }\href {\doibase 10.1103/PhysRevD.102.024029}
  {\bibfield  {journal} {\bibinfo  {journal} {Phys. Rev. D}\ }\textbf {\bibinfo
  {volume} {102}},\ \bibinfo {pages} {024029} (\bibinfo {year} {2020})},\
  \Eprint {http://arxiv.org/abs/2004.10716} {arXiv:2004.10716 [hep-th]}
  \BibitemShut {NoStop}%
\bibitem [{\citenamefont {Ai}(2020)}]{Ai:2020peo}%
  \BibitemOpen
  \bibfield  {author} {\bibinfo {author} {\bibfnamefont {W.-Y.}\ \bibnamefont
  {Ai}},\ }\href {\doibase 10.1088/1572-9494/aba242} {\bibfield  {journal}
  {\bibinfo  {journal} {Commun. Theor. Phys.}\ }\textbf {\bibinfo {volume}
  {72}},\ \bibinfo {pages} {095402} (\bibinfo {year} {2020})},\ \Eprint
  {http://arxiv.org/abs/2004.02858} {arXiv:2004.02858 [gr-qc]} \BibitemShut
  {NoStop}%
\bibitem [{\citenamefont {Wei}\ and\ \citenamefont {Liu}(2021)}]{Wei:2020ght}%
  \BibitemOpen
  \bibfield  {author} {\bibinfo {author} {\bibfnamefont {S.-W.}\ \bibnamefont
  {Wei}}\ and\ \bibinfo {author} {\bibfnamefont {Y.-X.}\ \bibnamefont {Liu}},\
  }\href {\doibase 10.1140/epjp/s13360-021-01398-9} {\bibfield  {journal}
  {\bibinfo  {journal} {Eur. Phys. J. Plus}\ }\textbf {\bibinfo {volume}
  {136}},\ \bibinfo {pages} {436} (\bibinfo {year} {2021})},\ \Eprint
  {http://arxiv.org/abs/2003.07769} {arXiv:2003.07769 [gr-qc]} \BibitemShut
  {NoStop}%
\bibitem [{\citenamefont {Aoki}\ \emph
  {et~al.}(2020{\natexlab{a}})\citenamefont {Aoki}, \citenamefont {Gorji},\
  and\ \citenamefont {Mukohyama}}]{Aoki:2020lig}%
  \BibitemOpen
  \bibfield  {author} {\bibinfo {author} {\bibfnamefont {K.}~\bibnamefont
  {Aoki}}, \bibinfo {author} {\bibfnamefont {M.~A.}\ \bibnamefont {Gorji}}, \
  and\ \bibinfo {author} {\bibfnamefont {S.}~\bibnamefont {Mukohyama}},\ }\href
  {\doibase 10.1016/j.physletb.2020.135843} {\bibfield  {journal} {\bibinfo
  {journal} {Phys. Lett. B}\ }\textbf {\bibinfo {volume} {810}},\ \bibinfo
  {pages} {135843} (\bibinfo {year} {2020}{\natexlab{a}})},\ \Eprint
  {http://arxiv.org/abs/2005.03859} {arXiv:2005.03859 [gr-qc]} \BibitemShut
  {NoStop}%
\bibitem [{\citenamefont {Nojiri}\ and\ \citenamefont
  {Odintsov}(2020)}]{Nojiri:2020tph}%
  \BibitemOpen
  \bibfield  {author} {\bibinfo {author} {\bibfnamefont {S.}~\bibnamefont
  {Nojiri}}\ and\ \bibinfo {author} {\bibfnamefont {S.~D.}\ \bibnamefont
  {Odintsov}},\ }\href {\doibase 10.1209/0295-5075/130/10004} {\bibfield
  {journal} {\bibinfo  {journal} {EPL}\ }\textbf {\bibinfo {volume} {130}},\
  \bibinfo {pages} {10004} (\bibinfo {year} {2020})},\ \Eprint
  {http://arxiv.org/abs/2004.01404} {arXiv:2004.01404 [hep-th]} \BibitemShut
  {NoStop}%
\bibitem [{\citenamefont {Konoplya}\ and\ \citenamefont
  {Zinhailo}(2020{\natexlab{a}})}]{Konoplya:2020bxa}%
  \BibitemOpen
  \bibfield  {author} {\bibinfo {author} {\bibfnamefont {R.~A.}\ \bibnamefont
  {Konoplya}}\ and\ \bibinfo {author} {\bibfnamefont {A.~F.}\ \bibnamefont
  {Zinhailo}},\ }\href {\doibase 10.1140/epjc/s10052-020-08639-8} {\bibfield
  {journal} {\bibinfo  {journal} {Eur. Phys. J. C}\ }\textbf {\bibinfo {volume}
  {80}},\ \bibinfo {pages} {1049} (\bibinfo {year} {2020}{\natexlab{a}})},\
  \Eprint {http://arxiv.org/abs/2003.01188} {arXiv:2003.01188 [gr-qc]}
  \BibitemShut {NoStop}%
\bibitem [{\citenamefont {Guo}\ and\ \citenamefont {Li}(2020)}]{Guo:2020zmf}%
  \BibitemOpen
  \bibfield  {author} {\bibinfo {author} {\bibfnamefont {M.}~\bibnamefont
  {Guo}}\ and\ \bibinfo {author} {\bibfnamefont {P.-C.}\ \bibnamefont {Li}},\
  }\href {\doibase 10.1140/epjc/s10052-020-8164-7} {\bibfield  {journal}
  {\bibinfo  {journal} {Eur. Phys. J. C}\ }\textbf {\bibinfo {volume} {80}},\
  \bibinfo {pages} {588} (\bibinfo {year} {2020})},\ \Eprint
  {http://arxiv.org/abs/2003.02523} {arXiv:2003.02523 [gr-qc]} \BibitemShut
  {NoStop}%
\bibitem [{\citenamefont {Fernandes}(2020)}]{Fernandes:2020rpa}%
  \BibitemOpen
  \bibfield  {author} {\bibinfo {author} {\bibfnamefont {P.~G.~S.}\
  \bibnamefont {Fernandes}},\ }\href {\doibase 10.1016/j.physletb.2020.135468}
  {\bibfield  {journal} {\bibinfo  {journal} {Phys. Lett. B}\ }\textbf
  {\bibinfo {volume} {805}},\ \bibinfo {pages} {135468} (\bibinfo {year}
  {2020})},\ \Eprint {http://arxiv.org/abs/2003.05491} {arXiv:2003.05491
  [gr-qc]} \BibitemShut {NoStop}%
\bibitem [{\citenamefont {Casalino}\ \emph {et~al.}(2021)\citenamefont
  {Casalino}, \citenamefont {Colleaux}, \citenamefont {Rinaldi},\ and\
  \citenamefont {Vicentini}}]{Casalino:2020kbt}%
  \BibitemOpen
  \bibfield  {author} {\bibinfo {author} {\bibfnamefont {A.}~\bibnamefont
  {Casalino}}, \bibinfo {author} {\bibfnamefont {A.}~\bibnamefont {Colleaux}},
  \bibinfo {author} {\bibfnamefont {M.}~\bibnamefont {Rinaldi}}, \ and\
  \bibinfo {author} {\bibfnamefont {S.}~\bibnamefont {Vicentini}},\ }\href
  {\doibase 10.1016/j.dark.2020.100770} {\bibfield  {journal} {\bibinfo
  {journal} {Phys. Dark Univ.}\ }\textbf {\bibinfo {volume} {31}},\ \bibinfo
  {pages} {100770} (\bibinfo {year} {2021})},\ \Eprint
  {http://arxiv.org/abs/2003.07068} {arXiv:2003.07068 [gr-qc]} \BibitemShut
  {NoStop}%
\bibitem [{\citenamefont {Hegde}\ \emph {et~al.}(2020)\citenamefont {Hegde},
  \citenamefont {Naveena~Kumara}, \citenamefont {Rizwan}, \citenamefont {M.},\
  and\ \citenamefont {Ali}}]{Hegde:2020xlv}%
  \BibitemOpen
  \bibfield  {author} {\bibinfo {author} {\bibfnamefont {K.}~\bibnamefont
  {Hegde}}, \bibinfo {author} {\bibfnamefont {A.}~\bibnamefont
  {Naveena~Kumara}}, \bibinfo {author} {\bibfnamefont {C.~L.~A.}\ \bibnamefont
  {Rizwan}}, \bibinfo {author} {\bibfnamefont {A.~K.}\ \bibnamefont {M.}}, \
  and\ \bibinfo {author} {\bibfnamefont {M.~S.}\ \bibnamefont {Ali}},\
  }\href@noop {} {\  (\bibinfo {year} {2020})},\ \Eprint
  {http://arxiv.org/abs/2003.08778} {arXiv:2003.08778 [gr-qc]} \BibitemShut
  {NoStop}%
\bibitem [{\citenamefont {Ghosh}\ and\ \citenamefont
  {Maharaj}(2020)}]{Ghosh:2020vpc}%
  \BibitemOpen
  \bibfield  {author} {\bibinfo {author} {\bibfnamefont {S.~G.}\ \bibnamefont
  {Ghosh}}\ and\ \bibinfo {author} {\bibfnamefont {S.~D.}\ \bibnamefont
  {Maharaj}},\ }\href {\doibase 10.1016/j.dark.2020.100687} {\bibfield
  {journal} {\bibinfo  {journal} {Phys. Dark Univ.}\ }\textbf {\bibinfo
  {volume} {30}},\ \bibinfo {pages} {100687} (\bibinfo {year} {2020})},\
  \Eprint {http://arxiv.org/abs/2003.09841} {arXiv:2003.09841 [gr-qc]}
  \BibitemShut {NoStop}%
\bibitem [{\citenamefont {Doneva}\ and\ \citenamefont
  {Yazadjiev}(2021)}]{Doneva:2020ped}%
  \BibitemOpen
  \bibfield  {author} {\bibinfo {author} {\bibfnamefont {D.~D.}\ \bibnamefont
  {Doneva}}\ and\ \bibinfo {author} {\bibfnamefont {S.~S.}\ \bibnamefont
  {Yazadjiev}},\ }\href {\doibase 10.1088/1475-7516/2021/05/024} {\bibfield
  {journal} {\bibinfo  {journal} {JCAP}\ }\textbf {\bibinfo {volume} {05}},\
  \bibinfo {pages} {024} (\bibinfo {year} {2021})},\ \Eprint
  {http://arxiv.org/abs/2003.10284} {arXiv:2003.10284 [gr-qc]} \BibitemShut
  {NoStop}%
\bibitem [{\citenamefont {Zhang}\ \emph
  {et~al.}(2020{\natexlab{a}})\citenamefont {Zhang}, \citenamefont {Wei},\ and\
  \citenamefont {Liu}}]{Zhang:2020qew}%
  \BibitemOpen
  \bibfield  {author} {\bibinfo {author} {\bibfnamefont {Y.-P.}\ \bibnamefont
  {Zhang}}, \bibinfo {author} {\bibfnamefont {S.-W.}\ \bibnamefont {Wei}}, \
  and\ \bibinfo {author} {\bibfnamefont {Y.-X.}\ \bibnamefont {Liu}},\ }\href
  {\doibase 10.3390/universe6080103} {\bibfield  {journal} {\bibinfo  {journal}
  {Universe}\ }\textbf {\bibinfo {volume} {6}},\ \bibinfo {pages} {103}
  (\bibinfo {year} {2020}{\natexlab{a}})},\ \Eprint
  {http://arxiv.org/abs/2003.10960} {arXiv:2003.10960 [gr-qc]} \BibitemShut
  {NoStop}%
\bibitem [{\citenamefont {Konoplya}\ and\ \citenamefont
  {Zhidenko}(2020{\natexlab{b}})}]{Konoplya:2020ibi}%
  \BibitemOpen
  \bibfield  {author} {\bibinfo {author} {\bibfnamefont {R.~A.}\ \bibnamefont
  {Konoplya}}\ and\ \bibinfo {author} {\bibfnamefont {A.}~\bibnamefont
  {Zhidenko}},\ }\href {\doibase 10.1103/PhysRevD.102.064004} {\bibfield
  {journal} {\bibinfo  {journal} {Phys. Rev. D}\ }\textbf {\bibinfo {volume}
  {102}},\ \bibinfo {pages} {064004} (\bibinfo {year} {2020}{\natexlab{b}})},\
  \Eprint {http://arxiv.org/abs/2003.12171} {arXiv:2003.12171 [gr-qc]}
  \BibitemShut {NoStop}%
\bibitem [{\citenamefont {Singh}\ and\ \citenamefont
  {Siwach}(2020)}]{Singh:2020xju}%
  \BibitemOpen
  \bibfield  {author} {\bibinfo {author} {\bibfnamefont {D.~V.}\ \bibnamefont
  {Singh}}\ and\ \bibinfo {author} {\bibfnamefont {S.}~\bibnamefont {Siwach}},\
  }\href {\doibase 10.1016/j.physletb.2020.135658} {\bibfield  {journal}
  {\bibinfo  {journal} {Phys. Lett. B}\ }\textbf {\bibinfo {volume} {808}},\
  \bibinfo {pages} {135658} (\bibinfo {year} {2020})},\ \Eprint
  {http://arxiv.org/abs/2003.11754} {arXiv:2003.11754 [gr-qc]} \BibitemShut
  {NoStop}%
\bibitem [{\citenamefont {Ghosh}\ and\ \citenamefont
  {Kumar}(2020)}]{Ghosh:2020syx}%
  \BibitemOpen
  \bibfield  {author} {\bibinfo {author} {\bibfnamefont {S.~G.}\ \bibnamefont
  {Ghosh}}\ and\ \bibinfo {author} {\bibfnamefont {R.}~\bibnamefont {Kumar}},\
  }\href {\doibase 10.1088/1361-6382/abc134} {\bibfield  {journal} {\bibinfo
  {journal} {Class. Quant. Grav.}\ }\textbf {\bibinfo {volume} {37}},\ \bibinfo
  {pages} {245008} (\bibinfo {year} {2020})},\ \Eprint
  {http://arxiv.org/abs/2003.12291} {arXiv:2003.12291 [gr-qc]} \BibitemShut
  {NoStop}%
\bibitem [{\citenamefont {Konoplya}\ and\ \citenamefont
  {Zhidenko}(2020{\natexlab{c}})}]{Konoplya:2020juj}%
  \BibitemOpen
  \bibfield  {author} {\bibinfo {author} {\bibfnamefont {R.~A.}\ \bibnamefont
  {Konoplya}}\ and\ \bibinfo {author} {\bibfnamefont {A.}~\bibnamefont
  {Zhidenko}},\ }\href {\doibase 10.1016/j.dark.2020.100697} {\bibfield
  {journal} {\bibinfo  {journal} {Phys. Dark Univ.}\ }\textbf {\bibinfo
  {volume} {30}},\ \bibinfo {pages} {100697} (\bibinfo {year}
  {2020}{\natexlab{c}})},\ \Eprint {http://arxiv.org/abs/2003.12492}
  {arXiv:2003.12492 [gr-qc]} \BibitemShut {NoStop}%
\bibitem [{\citenamefont {Kumar}\ and\ \citenamefont
  {Kumar}(2020)}]{Kumar:2020uyz}%
  \BibitemOpen
  \bibfield  {author} {\bibinfo {author} {\bibfnamefont {A.}~\bibnamefont
  {Kumar}}\ and\ \bibinfo {author} {\bibfnamefont {R.}~\bibnamefont {Kumar}},\
  }\href@noop {} {\  (\bibinfo {year} {2020})},\ \Eprint
  {http://arxiv.org/abs/2003.13104} {arXiv:2003.13104 [gr-qc]} \BibitemShut
  {NoStop}%
\bibitem [{\citenamefont {Zhang}\ \emph
  {et~al.}(2020{\natexlab{b}})\citenamefont {Zhang}, \citenamefont {Li},\ and\
  \citenamefont {Guo}}]{Zhang:2020qam}%
  \BibitemOpen
  \bibfield  {author} {\bibinfo {author} {\bibfnamefont {C.-Y.}\ \bibnamefont
  {Zhang}}, \bibinfo {author} {\bibfnamefont {P.-C.}\ \bibnamefont {Li}}, \
  and\ \bibinfo {author} {\bibfnamefont {M.}~\bibnamefont {Guo}},\ }\href
  {\doibase 10.1140/epjc/s10052-020-08448-z} {\bibfield  {journal} {\bibinfo
  {journal} {Eur. Phys. J. C}\ }\textbf {\bibinfo {volume} {80}},\ \bibinfo
  {pages} {874} (\bibinfo {year} {2020}{\natexlab{b}})},\ \Eprint
  {http://arxiv.org/abs/2003.13068} {arXiv:2003.13068 [hep-th]} \BibitemShut
  {NoStop}%
\bibitem [{\citenamefont {Hosseini~Mansoori}(2021)}]{HosseiniMansoori:2020yfj}%
  \BibitemOpen
  \bibfield  {author} {\bibinfo {author} {\bibfnamefont {S.~A.}\ \bibnamefont
  {Hosseini~Mansoori}},\ }\href {\doibase 10.1016/j.dark.2021.100776}
  {\bibfield  {journal} {\bibinfo  {journal} {Phys. Dark Univ.}\ }\textbf
  {\bibinfo {volume} {31}},\ \bibinfo {pages} {100776} (\bibinfo {year}
  {2021})},\ \Eprint {http://arxiv.org/abs/2003.13382} {arXiv:2003.13382
  [gr-qc]} \BibitemShut {NoStop}%
\bibitem [{\citenamefont {Wei}\ and\ \citenamefont {Liu}(2020)}]{Wei:2020poh}%
  \BibitemOpen
  \bibfield  {author} {\bibinfo {author} {\bibfnamefont {S.-W.}\ \bibnamefont
  {Wei}}\ and\ \bibinfo {author} {\bibfnamefont {Y.-X.}\ \bibnamefont {Liu}},\
  }\href {\doibase 10.1103/PhysRevD.101.104018} {\bibfield  {journal} {\bibinfo
   {journal} {Phys. Rev. D}\ }\textbf {\bibinfo {volume} {101}},\ \bibinfo
  {pages} {104018} (\bibinfo {year} {2020})},\ \Eprint
  {http://arxiv.org/abs/2003.14275} {arXiv:2003.14275 [gr-qc]} \BibitemShut
  {NoStop}%
\bibitem [{\citenamefont {Singh}\ \emph {et~al.}(2020)\citenamefont {Singh},
  \citenamefont {Ghosh},\ and\ \citenamefont {Maharaj}}]{Singh:2020nwo}%
  \BibitemOpen
  \bibfield  {author} {\bibinfo {author} {\bibfnamefont {D.~V.}\ \bibnamefont
  {Singh}}, \bibinfo {author} {\bibfnamefont {S.~G.}\ \bibnamefont {Ghosh}}, \
  and\ \bibinfo {author} {\bibfnamefont {S.~D.}\ \bibnamefont {Maharaj}},\
  }\href {\doibase 10.1016/j.dark.2020.100730} {\bibfield  {journal} {\bibinfo
  {journal} {Phys. Dark Univ.}\ }\textbf {\bibinfo {volume} {30}},\ \bibinfo
  {pages} {100730} (\bibinfo {year} {2020})},\ \Eprint
  {http://arxiv.org/abs/2003.14136} {arXiv:2003.14136 [gr-qc]} \BibitemShut
  {NoStop}%
\bibitem [{\citenamefont {Churilova}(2021)}]{Churilova:2020aca}%
  \BibitemOpen
  \bibfield  {author} {\bibinfo {author} {\bibfnamefont {M.~S.}\ \bibnamefont
  {Churilova}},\ }\href {\doibase 10.1016/j.dark.2020.100748} {\bibfield
  {journal} {\bibinfo  {journal} {Phys. Dark Univ.}\ }\textbf {\bibinfo
  {volume} {31}},\ \bibinfo {pages} {100748} (\bibinfo {year} {2021})},\
  \Eprint {http://arxiv.org/abs/2004.00513} {arXiv:2004.00513 [gr-qc]}
  \BibitemShut {NoStop}%
\bibitem [{\citenamefont {Islam}\ \emph {et~al.}(2020)\citenamefont {Islam},
  \citenamefont {Kumar},\ and\ \citenamefont {Ghosh}}]{Islam:2020xmy}%
  \BibitemOpen
  \bibfield  {author} {\bibinfo {author} {\bibfnamefont {S.~U.}\ \bibnamefont
  {Islam}}, \bibinfo {author} {\bibfnamefont {R.}~\bibnamefont {Kumar}}, \ and\
  \bibinfo {author} {\bibfnamefont {S.~G.}\ \bibnamefont {Ghosh}},\ }\href
  {\doibase 10.1088/1475-7516/2020/09/030} {\bibfield  {journal} {\bibinfo
  {journal} {JCAP}\ }\textbf {\bibinfo {volume} {09}},\ \bibinfo {pages} {030}
  (\bibinfo {year} {2020})},\ \Eprint {http://arxiv.org/abs/2004.01038}
  {arXiv:2004.01038 [gr-qc]} \BibitemShut {NoStop}%
\bibitem [{\citenamefont {Mishra}(2020)}]{Mishra:2020gce}%
  \BibitemOpen
  \bibfield  {author} {\bibinfo {author} {\bibfnamefont {A.~K.}\ \bibnamefont
  {Mishra}},\ }\href {\doibase 10.1007/s10714-020-02763-2} {\bibfield
  {journal} {\bibinfo  {journal} {Gen. Rel. Grav.}\ }\textbf {\bibinfo {volume}
  {52}},\ \bibinfo {pages} {106} (\bibinfo {year} {2020})},\ \Eprint
  {http://arxiv.org/abs/2004.01243} {arXiv:2004.01243 [gr-qc]} \BibitemShut
  {NoStop}%
\bibitem [{\citenamefont {Konoplya}\ and\ \citenamefont
  {Zinhailo}(2020{\natexlab{b}})}]{Konoplya:2020cbv}%
  \BibitemOpen
  \bibfield  {author} {\bibinfo {author} {\bibfnamefont {R.~A.}\ \bibnamefont
  {Konoplya}}\ and\ \bibinfo {author} {\bibfnamefont {A.~F.}\ \bibnamefont
  {Zinhailo}},\ }\href {\doibase 10.1016/j.physletb.2020.135793} {\bibfield
  {journal} {\bibinfo  {journal} {Phys. Lett. B}\ }\textbf {\bibinfo {volume}
  {810}},\ \bibinfo {pages} {135793} (\bibinfo {year} {2020}{\natexlab{b}})},\
  \Eprint {http://arxiv.org/abs/2004.02248} {arXiv:2004.02248 [gr-qc]}
  \BibitemShut {NoStop}%
\bibitem [{\citenamefont {Zhang}\ \emph
  {et~al.}(2020{\natexlab{c}})\citenamefont {Zhang}, \citenamefont {Zhang},
  \citenamefont {Li},\ and\ \citenamefont {Guo}}]{Zhang:2020sjh}%
  \BibitemOpen
  \bibfield  {author} {\bibinfo {author} {\bibfnamefont {C.-Y.}\ \bibnamefont
  {Zhang}}, \bibinfo {author} {\bibfnamefont {S.-J.}\ \bibnamefont {Zhang}},
  \bibinfo {author} {\bibfnamefont {P.-C.}\ \bibnamefont {Li}}, \ and\ \bibinfo
  {author} {\bibfnamefont {M.}~\bibnamefont {Guo}},\ }\href {\doibase
  10.1007/JHEP08(2020)105} {\bibfield  {journal} {\bibinfo  {journal} {JHEP}\
  }\textbf {\bibinfo {volume} {08}},\ \bibinfo {pages} {105} (\bibinfo {year}
  {2020}{\natexlab{c}})},\ \Eprint {http://arxiv.org/abs/2004.03141}
  {arXiv:2004.03141 [gr-qc]} \BibitemShut {NoStop}%
\bibitem [{\citenamefont {Eslam~Panah}\ \emph {et~al.}(2020)\citenamefont
  {Eslam~Panah}, \citenamefont {Jafarzade},\ and\ \citenamefont
  {Hendi}}]{EslamPanah:2020hoj}%
  \BibitemOpen
  \bibfield  {author} {\bibinfo {author} {\bibfnamefont {B.}~\bibnamefont
  {Eslam~Panah}}, \bibinfo {author} {\bibfnamefont {K.}~\bibnamefont
  {Jafarzade}}, \ and\ \bibinfo {author} {\bibfnamefont {S.~H.}\ \bibnamefont
  {Hendi}},\ }\href {\doibase 10.1016/j.nuclphysb.2020.115269} {\bibfield
  {journal} {\bibinfo  {journal} {Nucl. Phys. B}\ }\textbf {\bibinfo {volume}
  {961}},\ \bibinfo {pages} {115269} (\bibinfo {year} {2020})},\ \Eprint
  {http://arxiv.org/abs/2004.04058} {arXiv:2004.04058 [hep-th]} \BibitemShut
  {NoStop}%
\bibitem [{\citenamefont {Arag\'on}\ \emph {et~al.}(2020)\citenamefont
  {Arag\'on}, \citenamefont {B\'ecar}, \citenamefont {Gonz\'alez},\ and\
  \citenamefont {V\'asquez}}]{Aragon:2020qdc}%
  \BibitemOpen
  \bibfield  {author} {\bibinfo {author} {\bibfnamefont {A.}~\bibnamefont
  {Arag\'on}}, \bibinfo {author} {\bibfnamefont {R.}~\bibnamefont {B\'ecar}},
  \bibinfo {author} {\bibfnamefont {P.~A.}\ \bibnamefont {Gonz\'alez}}, \ and\
  \bibinfo {author} {\bibfnamefont {Y.}~\bibnamefont {V\'asquez}},\ }\href
  {\doibase 10.1140/epjc/s10052-020-8298-7} {\bibfield  {journal} {\bibinfo
  {journal} {Eur. Phys. J. C}\ }\textbf {\bibinfo {volume} {80}},\ \bibinfo
  {pages} {773} (\bibinfo {year} {2020})},\ \Eprint
  {http://arxiv.org/abs/2004.05632} {arXiv:2004.05632 [gr-qc]} \BibitemShut
  {NoStop}%
\bibitem [{\citenamefont {Aoki}\ \emph
  {et~al.}(2020{\natexlab{b}})\citenamefont {Aoki}, \citenamefont {Gorji},\
  and\ \citenamefont {Mukohyama}}]{Aoki:2020iwm}%
  \BibitemOpen
  \bibfield  {author} {\bibinfo {author} {\bibfnamefont {K.}~\bibnamefont
  {Aoki}}, \bibinfo {author} {\bibfnamefont {M.~A.}\ \bibnamefont {Gorji}}, \
  and\ \bibinfo {author} {\bibfnamefont {S.}~\bibnamefont {Mukohyama}},\ }\href
  {\doibase 10.1088/1475-7516/2020/09/014} {\bibfield  {journal} {\bibinfo
  {journal} {JCAP}\ }\textbf {\bibinfo {volume} {09}},\ \bibinfo {pages} {014}
  (\bibinfo {year} {2020}{\natexlab{b}})},\ \bibinfo {note} {[Erratum: JCAP 05,
  E01 (2021)]},\ \Eprint {http://arxiv.org/abs/2005.08428} {arXiv:2005.08428
  [gr-qc]} \BibitemShut {NoStop}%
\bibitem [{\citenamefont {Shu}(2020)}]{Shu:2020cjw}%
  \BibitemOpen
  \bibfield  {author} {\bibinfo {author} {\bibfnamefont {F.-W.}\ \bibnamefont
  {Shu}},\ }\href {\doibase 10.1016/j.physletb.2020.135907} {\bibfield
  {journal} {\bibinfo  {journal} {Phys. Lett. B}\ }\textbf {\bibinfo {volume}
  {811}},\ \bibinfo {pages} {135907} (\bibinfo {year} {2020})},\ \Eprint
  {http://arxiv.org/abs/2004.09339} {arXiv:2004.09339 [gr-qc]} \BibitemShut
  {NoStop}%
\bibitem [{\citenamefont {Mahapatra}(2020)}]{Mahapatra:2020rds}%
  \BibitemOpen
  \bibfield  {author} {\bibinfo {author} {\bibfnamefont {S.}~\bibnamefont
  {Mahapatra}},\ }\href {\doibase 10.1140/epjc/s10052-020-08568-6} {\bibfield
  {journal} {\bibinfo  {journal} {Eur. Phys. J. C}\ }\textbf {\bibinfo {volume}
  {80}},\ \bibinfo {pages} {992} (\bibinfo {year} {2020})},\ \Eprint
  {http://arxiv.org/abs/2004.09214} {arXiv:2004.09214 [gr-qc]} \BibitemShut
  {NoStop}%
\bibitem [{\citenamefont {Lu}\ and\ \citenamefont {Pang}(2020)}]{Lu:2020iav}%
  \BibitemOpen
  \bibfield  {author} {\bibinfo {author} {\bibfnamefont {H.}~\bibnamefont
  {Lu}}\ and\ \bibinfo {author} {\bibfnamefont {Y.}~\bibnamefont {Pang}},\
  }\href {\doibase 10.1016/j.physletb.2020.135717} {\bibfield  {journal}
  {\bibinfo  {journal} {Phys. Lett. B}\ }\textbf {\bibinfo {volume} {809}},\
  \bibinfo {pages} {135717} (\bibinfo {year} {2020})},\ \Eprint
  {http://arxiv.org/abs/2003.11552} {arXiv:2003.11552 [gr-qc]} \BibitemShut
  {NoStop}%
\bibitem [{\citenamefont {G\"urses}\ \emph {et~al.}(2020)\citenamefont
  {G\"urses}, \citenamefont {\c{S}i\c{s}man},\ and\ \citenamefont
  {Tekin}}]{Gurses:2020ofy}%
  \BibitemOpen
  \bibfield  {author} {\bibinfo {author} {\bibfnamefont {M.}~\bibnamefont
  {G\"urses}}, \bibinfo {author} {\bibfnamefont {T.~c.}\ \bibnamefont
  {\c{S}i\c{s}man}}, \ and\ \bibinfo {author} {\bibfnamefont {B.}~\bibnamefont
  {Tekin}},\ }\href {\doibase 10.1140/epjc/s10052-020-8200-7} {\bibfield
  {journal} {\bibinfo  {journal} {Eur. Phys. J. C}\ }\textbf {\bibinfo {volume}
  {80}},\ \bibinfo {pages} {647} (\bibinfo {year} {2020})},\ \Eprint
  {http://arxiv.org/abs/2004.03390} {arXiv:2004.03390 [gr-qc]} \BibitemShut
  {NoStop}%
\bibitem [{\citenamefont {Banerjee}\ \emph {et~al.}(2021)\citenamefont
  {Banerjee}, \citenamefont {Tangphati}, \citenamefont {Samart},\ and\
  \citenamefont {Channuie}}]{Banerjee:2020dad}%
  \BibitemOpen
  \bibfield  {author} {\bibinfo {author} {\bibfnamefont {A.}~\bibnamefont
  {Banerjee}}, \bibinfo {author} {\bibfnamefont {T.}~\bibnamefont {Tangphati}},
  \bibinfo {author} {\bibfnamefont {D.}~\bibnamefont {Samart}}, \ and\ \bibinfo
  {author} {\bibfnamefont {P.}~\bibnamefont {Channuie}},\ }\href {\doibase
  10.3847/1538-4357/abc87f} {\bibfield  {journal} {\bibinfo  {journal}
  {Astrophys. J.}\ }\textbf {\bibinfo {volume} {906}},\ \bibinfo {pages} {114}
  (\bibinfo {year} {2021})},\ \Eprint {http://arxiv.org/abs/2007.04121}
  {arXiv:2007.04121 [gr-qc]} \BibitemShut {NoStop}%
\bibitem [{\citenamefont {Ge}\ and\ \citenamefont {Sin}(2020)}]{Ge:2020tid}%
  \BibitemOpen
  \bibfield  {author} {\bibinfo {author} {\bibfnamefont {X.-H.}\ \bibnamefont
  {Ge}}\ and\ \bibinfo {author} {\bibfnamefont {S.-J.}\ \bibnamefont {Sin}},\
  }\href {\doibase 10.1140/epjc/s10052-020-8288-9} {\bibfield  {journal}
  {\bibinfo  {journal} {Eur. Phys. J. C}\ }\textbf {\bibinfo {volume} {80}},\
  \bibinfo {pages} {695} (\bibinfo {year} {2020})},\ \Eprint
  {http://arxiv.org/abs/2004.12191} {arXiv:2004.12191 [hep-th]} \BibitemShut
  {NoStop}%
\bibitem [{\citenamefont {Yang}\ \emph
  {et~al.}(2020{\natexlab{a}})\citenamefont {Yang}, \citenamefont {Gu},
  \citenamefont {Wei},\ and\ \citenamefont {Liu}}]{Yang:2020jno}%
  \BibitemOpen
  \bibfield  {author} {\bibinfo {author} {\bibfnamefont {K.}~\bibnamefont
  {Yang}}, \bibinfo {author} {\bibfnamefont {B.-M.}\ \bibnamefont {Gu}},
  \bibinfo {author} {\bibfnamefont {S.-W.}\ \bibnamefont {Wei}}, \ and\
  \bibinfo {author} {\bibfnamefont {Y.-X.}\ \bibnamefont {Liu}},\ }\href
  {\doibase 10.1140/epjc/s10052-020-8246-6} {\bibfield  {journal} {\bibinfo
  {journal} {Eur. Phys. J. C}\ }\textbf {\bibinfo {volume} {80}},\ \bibinfo
  {pages} {662} (\bibinfo {year} {2020}{\natexlab{a}})},\ \Eprint
  {http://arxiv.org/abs/2004.14468} {arXiv:2004.14468 [gr-qc]} \BibitemShut
  {NoStop}%
\bibitem [{\citenamefont {Lin}\ \emph {et~al.}(2020)\citenamefont {Lin},
  \citenamefont {Yang}, \citenamefont {Wei}, \citenamefont {Wang},\ and\
  \citenamefont {Liu}}]{Lin:2020kqe}%
  \BibitemOpen
  \bibfield  {author} {\bibinfo {author} {\bibfnamefont {Z.-C.}\ \bibnamefont
  {Lin}}, \bibinfo {author} {\bibfnamefont {K.}~\bibnamefont {Yang}}, \bibinfo
  {author} {\bibfnamefont {S.-W.}\ \bibnamefont {Wei}}, \bibinfo {author}
  {\bibfnamefont {Y.-Q.}\ \bibnamefont {Wang}}, \ and\ \bibinfo {author}
  {\bibfnamefont {Y.-X.}\ \bibnamefont {Liu}},\ }\href {\doibase
  10.1140/epjc/s10052-020-08612-5} {\bibfield  {journal} {\bibinfo  {journal}
  {Eur. Phys. J. C}\ }\textbf {\bibinfo {volume} {80}},\ \bibinfo {pages}
  {1033} (\bibinfo {year} {2020})},\ \Eprint {http://arxiv.org/abs/2006.07913}
  {arXiv:2006.07913 [gr-qc]} \BibitemShut {NoStop}%
\bibitem [{\citenamefont {Yang}\ \emph
  {et~al.}(2020{\natexlab{b}})\citenamefont {Yang}, \citenamefont {Wan},
  \citenamefont {Chen}, \citenamefont {Yang},\ and\ \citenamefont
  {Wang}}]{Yang:2020czk}%
  \BibitemOpen
  \bibfield  {author} {\bibinfo {author} {\bibfnamefont {S.-J.}\ \bibnamefont
  {Yang}}, \bibinfo {author} {\bibfnamefont {J.-J.}\ \bibnamefont {Wan}},
  \bibinfo {author} {\bibfnamefont {J.}~\bibnamefont {Chen}}, \bibinfo {author}
  {\bibfnamefont {J.}~\bibnamefont {Yang}}, \ and\ \bibinfo {author}
  {\bibfnamefont {Y.-Q.}\ \bibnamefont {Wang}},\ }\href {\doibase
  10.1140/epjc/s10052-020-08511-9} {\bibfield  {journal} {\bibinfo  {journal}
  {Eur. Phys. J. C}\ }\textbf {\bibinfo {volume} {80}},\ \bibinfo {pages} {937}
  (\bibinfo {year} {2020}{\natexlab{b}})},\ \Eprint
  {http://arxiv.org/abs/2004.07934} {arXiv:2004.07934 [gr-qc]} \BibitemShut
  {NoStop}%
\bibitem [{\citenamefont {Gurses}\ \emph {et~al.}(2020)\citenamefont {Gurses},
  \citenamefont {{S}i{s}man},\ and\ \citenamefont {Tekin}}]{Gurses:2020rxb}%
  \BibitemOpen
  \bibfield  {author} {\bibinfo {author} {\bibfnamefont {M.}~\bibnamefont
  {Gurses}}, \bibinfo {author} {\bibfnamefont {T.~C.}\ \bibnamefont
  {{S}i{s}man}}, \ and\ \bibinfo {author} {\bibfnamefont {B.}~\bibnamefont
  {Tekin}},\ }\href {\doibase 10.1103/PhysRevLett.125.149001} {\bibfield
  {journal} {\bibinfo  {journal} {Phys. Rev. Lett.}\ }\textbf {\bibinfo
  {volume} {125}},\ \bibinfo {pages} {149001} (\bibinfo {year} {2020})},\
  \Eprint {http://arxiv.org/abs/2009.13508} {arXiv:2009.13508 [gr-qc]}
  \BibitemShut {NoStop}%
\bibitem [{\citenamefont {Corian\`o}\ \emph {et~al.}(2021)\citenamefont
  {Corian\`o}, \citenamefont {Maglio},\ and\ \citenamefont
  {Theofilopoulos}}]{Coriano:2021nvn}%
  \BibitemOpen
  \bibfield  {author} {\bibinfo {author} {\bibfnamefont {C.}~\bibnamefont
  {Corian\`o}}, \bibinfo {author} {\bibfnamefont {M.~M.}\ \bibnamefont
  {Maglio}}, \ and\ \bibinfo {author} {\bibfnamefont {D.}~\bibnamefont
  {Theofilopoulos}},\ }\href {\doibase 10.1140/epjc/s10052-021-09523-9} {\
  (\bibinfo {year} {2021}),\ 10.1140/epjc/s10052-021-09523-9},\ \Eprint
  {http://arxiv.org/abs/2103.13957} {arXiv:2103.13957 [hep-th]} \BibitemShut
  {NoStop}%
\bibitem [{\citenamefont {Corian\`o}\ \emph {et~al.}()\citenamefont
  {Corian\`o}, \citenamefont {Maglio},\ and\ \citenamefont
  {Theofilopoulos}}]{CMT}%
  \BibitemOpen
  \bibfield  {author} {\bibinfo {author} {\bibfnamefont {C.}~\bibnamefont
  {Corian\`o}}, \bibinfo {author} {\bibfnamefont {M.~M.}\ \bibnamefont
  {Maglio}}, \ and\ \bibinfo {author} {\bibfnamefont {D.}~\bibnamefont
  {Theofilopoulos}},\ }\href@noop {} {\ }\Eprint {http://arxiv.org/abs/in
  preparation} {in preparation} \BibitemShut {NoStop}%
\bibitem [{\citenamefont {'t~Hooft}(2016)}]{tHooft:2016uxd}%
  \BibitemOpen
  \bibfield  {author} {\bibinfo {author} {\bibfnamefont {G.}~\bibnamefont
  {'t~Hooft}},\ }\href {\doibase 10.1142/S0218271817300063} {\bibfield
  {journal} {\bibinfo  {journal} {Int. J. Mod. Phys. D}\ }\textbf {\bibinfo
  {volume} {26}},\ \bibinfo {pages} {1730006} (\bibinfo {year}
  {2016})}\BibitemShut {NoStop}%
\bibitem [{\citenamefont {'t~Hooft}(2015)}]{tHooft:2015koi}%
  \BibitemOpen
  \bibfield  {author} {\bibinfo {author} {\bibfnamefont {G.}~\bibnamefont
  {'t~Hooft}},\ }\href {\doibase 10.1093/acprof:oso/9780198727965.003.0010}
  {\bibfield  {journal} {\bibinfo  {journal} {Les Houches Lect. Notes}\
  }\textbf {\bibinfo {volume} {97}},\ \bibinfo {pages} {209} (\bibinfo {year}
  {2015})}\BibitemShut {NoStop}%
\bibitem [{\citenamefont {Corian\`o}\ \emph {et~al.}(2006)\citenamefont
  {Corian\`o}, \citenamefont {Irges},\ and\ \citenamefont
  {Kiritsis}}]{Coriano:2005js}%
  \BibitemOpen
  \bibfield  {author} {\bibinfo {author} {\bibfnamefont {C.}~\bibnamefont
  {Corian\`o}}, \bibinfo {author} {\bibfnamefont {N.}~\bibnamefont {Irges}}, \
  and\ \bibinfo {author} {\bibfnamefont {E.}~\bibnamefont {Kiritsis}},\ }\href
  {\doibase 10.1016/j.nuclphysb.2006.04.009} {\bibfield  {journal} {\bibinfo
  {journal} {Nucl. Phys.}\ }\textbf {\bibinfo {volume} {B746}},\ \bibinfo
  {pages} {77} (\bibinfo {year} {2006})},\ \Eprint
  {http://arxiv.org/abs/hep-ph/0510332} {arXiv:hep-ph/0510332} \BibitemShut
  {NoStop}%
\bibitem [{\citenamefont {Corian\`o}\ \emph {et~al.}(2019)\citenamefont
  {Corian\`o}, \citenamefont {Maglio},\ and\ \citenamefont
  {Mottola}}]{Coriano:2017mux}%
  \BibitemOpen
  \bibfield  {author} {\bibinfo {author} {\bibfnamefont {C.}~\bibnamefont
  {Corian\`o}}, \bibinfo {author} {\bibfnamefont {M.~M.}\ \bibnamefont
  {Maglio}}, \ and\ \bibinfo {author} {\bibfnamefont {E.}~\bibnamefont
  {Mottola}},\ }\href {\doibase 10.1016/j.nuclphysb.2019.03.019} {\bibfield
  {journal} {\bibinfo  {journal} {Nucl. Phys.}\ }\textbf {\bibinfo {volume}
  {B942}},\ \bibinfo {pages} {303} (\bibinfo {year} {2019})},\ \Eprint
  {http://arxiv.org/abs/1703.08860} {arXiv:1703.08860 [hep-th]} \BibitemShut
  {NoStop}%
\bibitem [{\citenamefont {Mazur}\ and\ \citenamefont
  {Mottola}(2001)}]{Mazur:2001aa}%
  \BibitemOpen
  \bibfield  {author} {\bibinfo {author} {\bibfnamefont {P.~O.}\ \bibnamefont
  {Mazur}}\ and\ \bibinfo {author} {\bibfnamefont {E.}~\bibnamefont
  {Mottola}},\ }\href {\doibase 10.1103/PhysRevD.64.104022} {\bibfield
  {journal} {\bibinfo  {journal} {Phys.Rev.}\ }\textbf {\bibinfo {volume}
  {D64}},\ \bibinfo {pages} {104022} (\bibinfo {year} {2001})},\ \Eprint
  {http://arxiv.org/abs/hep-th/0106151} {arXiv:hep-th/0106151 [hep-th]}
  \BibitemShut {NoStop}%
\bibitem [{\citenamefont {Riegert}(1984)}]{Riegert:1987kt}%
  \BibitemOpen
  \bibfield  {author} {\bibinfo {author} {\bibfnamefont {R.~J.}\ \bibnamefont
  {Riegert}},\ }\href {\doibase http://dx.doi.org/10.1016/0370-2693(84)90983-3}
  {\bibfield  {journal} {\bibinfo  {journal} {Phys.Lett.}\ }\textbf {\bibinfo
  {volume} {B134}},\ \bibinfo {pages} {56} (\bibinfo {year}
  {1984})}\BibitemShut {NoStop}%
\end{thebibliography}

 \end{document}